\documentclass[twocolumn,showpacs,preprintnumbers,amsmath,amssymb,epsfig,widetext]{revtex4}

\usepackage{graphicx}
\usepackage{dcolumn}
\usepackage{bm}
\usepackage{epsfig}

\def\fun#1#2{\lower3.6pt\vbox{\baselineskip0pt\lineskip.9pt
  \ialign{$\mathsurround=0pt#1\hfil##\hfil$\crcr#2\crcr\sim\crcr}}}

\input epsf

\def\be{\begin{equation}}
\def\ee{\end{equation}}
\def\ba{\begin{eqnarray}}
\def\ea{\end{eqnarray}}

\def\nn{\nonumber}

\begin{document}

\preprint{}

\title{Detectability of a phantom-like braneworld model\\ with the integrated 
Sachs-Wolfe effect}

\author{Tommaso Giannantonio \footnote{Tommaso.Giannantonio@port.ac.uk}, 
Yong-Seon Song \footnote{Yong-seon.Song@port.ac.uk} and Kazuya Koyama \footnote{Kazuya.Koyama@port.ac.uk}}
\affiliation{Institute of Cosmology \& Gravitation, 
University of Portsmouth, Portsmouth, Hampshire, PO1 2EG, UK }

\date{\today}

\begin{abstract}
We study a braneworld model in which a phantom-like behaviour 
occurs with only cold dark matter and a cosmological constant, 
due to a large distance modification of gravity. With the addition 
of curvature, the geometrical tests are not strict enough to rule out  
models in which gravity is modified significantly on large scales. 
We show that this degeneracy in the parameter space 
is broken by the structure formation tests, such as the integrated 
Sachs-Wolfe effect, which can probe general 
relativity on large scales.
\end{abstract}



\maketitle

\section{introduction}
The late time acceleration of the Universe is one of the 
biggest problems in cosmology. In the framework of conventional 
general relativity, the expansion of the Universe at late times is dominated 
by a \emph{dark energy} with negative pressure and equation of state $w \equiv p / \rho < -1/3$. 
Several current observations suggest $w<-1$, which is often called 
\emph{phantom} dark energy, although the fiducial LCDM model with $ w = -1 $ is still preferred 
if we combine all the data sets \cite{Percival:2007yw,Giannantonio:2008zi}. 
From a theoretical point of view, it is extremely difficult to realise dark energy 
models with $w<-1$: 
the easiest way to obtain such a model is to consider a ghost scalar field 
with the wrong sign for the kinetic term, although this leads to the instability 
of the vacuum \cite{Caldwell:1999ew}. There are a few successful models 
that lead to $w<-1$ without having theoretical pathologies \cite{Csaki:2005vq,
Libanov:2007mq}; 
among them, we focus on a braneworld model proposed by Sahni and Shtanov 
\cite{Sahni:2002dx}
and further developed by Lue and Starkman \cite{Lue:2004za}.

This model is based on the Dvali--Gabadadze--Porrati (DGP)
model of the 5D braneworld where we are supposed to live on 
a 4D brane in the 5D Minkowski spacetime \cite{Dvali:2000hr}. 
The 4D gravity on the brane is recovered by the induced 4D Einstein--Hilbert 
action on the brane. In this model there are two branches of the solutions
\cite{Deffayet:2000uy}:
in the first branch, known as \emph{self-accelerating}, the 
late time acceleration can be realised without introducing any
dark energy, while in the other, known as the \emph{normal} 
branch, a cosmological constant is needed to explain 
the late time accelerated expansion of the Universe; nevertheless, 
the extra-dimensional effects modify gravity on large scales 
and the model deviates from the standard LCDM. 
In particular, at the background level, the Universe behaves as if 
there were a phantom-like dark energy $w<-1$. 

Besides the fact that this model mimics a phantom behaviour,
it is known to be free of ghosts and thus represents a 
healthy modified gravity theory. This is in contrast with 
the self-accelerating branch of the DGP model (hereafter sDGP) 
where there exists 
a ghost at the linearised level (for a review see \cite{Koyama:2007za}). 
Another advantage of the model
is that there is a mechanism to recover general relativity 
on small scales. Thus with this model we can modify gravity 
on large scales significantly without spoiling the success of 
general relativity on the solar system scales, providing the 
basis for the test of the large distance modification of general relativity. 

In this paper, we study the phenomenological consequences of the 
normal branch DGP model (hereafter nDGP). 
We first present in Section II the geometrical 
tests on nDGP, looking for a parameter space 
which can be tested from structure formation, which is summarised in Section III.
Then we present the ISW-galaxy correlations as a powerful tool to distinguish 
between LCDM and nDGP models in Section IV. 
Section V is devoted to the conclusion. 

\section{Geometrical tests}
The cosmic expansion of the nDGP model depends on the usual 4D FRW metric plus
 the gravitational effect of the 5D bulk on the brane.
The cosmic acceleration is then introduced by the brane tension, which works
as a cosmological constant on the brane.
The gravity at large scales is modified by the 5D gravity effects on the brane,
which are parameterised by a transition scale from 4D gravity to 5D gravity.
The crossover distance $r_c$ is defined as the ratio between 4D and 5D Planck
mass scales
\ba
r_c=\frac{M_{4}^{2}}{2M_{5}^{3}}\,,
\ea
where $M_4$ and $M_5$ are the Planck scales in the 4D and 5D spacetime 
respectively. The late time expansion history is determined by two 
free parameters, the cosmological constant (or brane tension) $\Lambda$ and the crossover 
distance $r_c$.

The Friedmann equation for an nDGP model with curvature $ K = - \Omega_k H_0^2 $ is given by
\ba\label{eq:FRW}
H^2-\frac{\Omega_k H_0^2}{a^2}+\frac{1}{r_c}
\sqrt{H^2 -\frac{\Omega_k H_0^2}{a^2}} 
=\frac{8\pi G}{3}\rho_m+\frac{\Lambda}{3}\,,
\ea
and the dimensionless expansion history $E(a)$ is defined by
\ba
E^2(a)\equiv\frac{H^2(a)}{H_0^2}=\frac{\Omega_k}{a^2}+
\left[ \sqrt{\frac{\Omega_m}{a^3}+\Omega_{\Lambda}+\Omega_{r_c}}
-\sqrt{\Omega_{r_c}}\right]^2\,,
\ea
where $\Omega_{\Lambda}=\Lambda/3H_0^2$ and $\Omega_{r_c}=(4H_0^2r_c^2)^{-1}$,
which satisfies
\ba
\sqrt{\Omega_{r_c}}=\frac{\Omega_m+\Omega_{\Lambda}+\Omega_k-1}
{2\sqrt{1-\Omega_k}}\,.
\ea
The free parameter $r_c$ can range in theory from $0$ to the infinity;
however, it has been shown that the deviations from general 
relativity on solar system scales are also controlled by $r_c$,
and the current constraints require that $r_c > H_0^{-1}$. 
We can see that if $r_c$ approaches the infinity, then Eq.~(\ref{eq:FRW}) converges
to GR, while if $r_c$ approaches $H_0^{-1}$, then the 5D gravitational effect on
the expansion history becomes maximal.

The modification of gravity at late time screens the 
cosmological constant and makes the effective equation
of state less than $-1$. We define the effective 
energy density of dark energy $\rho_{\rm eff}$ as \cite{Lazkoz:2006gp}
\ba
H^2-\frac{\Omega_k H_0^2}{a^2}&=&
\frac{8\pi G}{3}\rho_m+\frac{8\pi G}{3}\rho_{\rm eff} \nn\\
\rho_{\rm eff}&=&\frac{1}{8\pi G}\left(\Lambda-\frac{3}{r_c}
\sqrt{H^2-\frac{\Omega_k H_0^2}{a^2}} \right)\,.
\ea
It is clearly seen that the 5D effects make the effective 
dark energy density $\rho_{\rm eff}$ smaller. 
From the continuity equation of $\rho_{\rm eff}$
\ba
\dot\rho_{\rm eff}+3H(1+w_{\rm eff})\rho_{\rm eff}=0\,,
\ea
we can derive $w_{\rm eff}$ as
\ba\label{eq:weff}
w_{\rm eff}&=&-1
-\frac{\sqrt{\Omega_{r_c}} \Omega_m a^{-3}}
{ \Omega_{\Lambda} - 2 \sqrt{\Omega_{r_c}}(E^2-\Omega_k/a^2)^{1/2}}\nonumber\\
&& \times \frac{1}{    
(E^2-\Omega_k/a^2)^{1/2} + \sqrt{\Omega_{r_c}}}
\,.
\ea
At the current time, the effective equation of state becomes 
\be
w_{\rm eff}(a=1)=-1 - \frac{(\Omega_m+\Omega_{\Lambda}-1)\Omega_m}
{(1-\Omega_m)(\Omega_m +\Omega_{\Lambda}+1)},
\ee
where we neglected the curvature for simplicity. Provided that 
$\Omega_m <1$, we have the phantom behaviour $w_{\rm eff}<-1$. 

\begin{figure}[htbp]
  \begin{center}
  \epsfysize=3.0truein
  \epsfxsize=3.0truein
    \epsffile{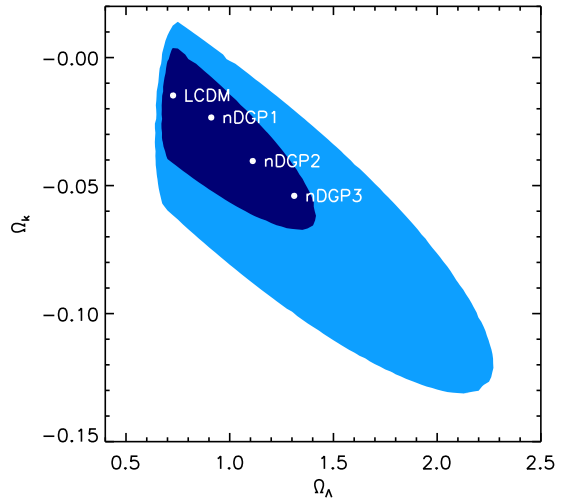}
    \caption{\footnotesize Geometrical test on the nDGP by using 
SN$+$CMB$+$$H_0$ observations. There is correlation observed 
in the projected $\Omega_k$ and $\Omega_{\Lambda}$ plane
after marginalisation of all other cosmological parameters.
}
\label{fig:contour}
\end{center}
\end{figure} 

We revisit the geometrical test on the nDGP \cite{Lazkoz:2006gp, Lazkoz:2007zk}.
The geometrical test on the nDGP with a flat curvature prior
is not in favour of the cases for the significant screening effect,
which rules out observable modified gravity effects in the nDGP.
However we find that measurable screening effects are 
allowed with the inclusion of curvature. We exploit the leverage arm in the 
geometrical tests at both ends of low and high redshifts. 
At low redshifts, we use the Gold SN data set \cite{riess04}. 
At high redshifts, we fix the distance to the last scattering surface at 
$z_{\mathrm{lss}}=1088^{+1}_{-2}$ by fitting the harmonic space scale of the 
acoustic peak $l^{A}=302^{+0.9}_{-1.4}$ and matter density 
$\Omega_mh^2=0.1268^{+0.0072}_{-0.0095}$ \cite{spergel03}.
In addition to that, we constrain the expansion constant $H_0$
with the Hubble constant measurement, 
$H_0=72^{+8}_{-8}$ \cite{Freedman:2000cf}.

With a fixed CMB prior of $\Omega_mh^2$, best fit values for $w$ and $H_0$
are correlated with each other.
The theoretical models predicting $w<-1$ have a smaller best fit value 
for $H_0$ compared with LCDM ($w=-1$).
Since the measured comoving distance to $z_{\mathrm{lss}}$ is consistent with
a best fit value for $H_0$ in flat LCDM,
the comoving distance to $z_{\mathrm{lss}}$ in phantom-like braneworld models 
becomes longer than the measured distance.
This worse fit for the large distance measured by CMB in the models with $w<-1$
can be cured by introducing a positive curvature
which makes the distance shorter without significantly affecting 
the fit for the shorter distance measured by SNe.
Consequently, a larger $\Omega_{\Lambda}$, which realises larger 
screening effects and $w<-1$, is allowed with a positive
curvature ($\Omega_k <0$) as is shown in Fig.~\ref{fig:contour}.
Hence if the curvature is added, there appears a degeneracy in the 
geometrical tests and the models with large modified gravity effects
are allowed. This degeneracy can be broken by the structure formation 
test.

\section{Structure formation tests}
There are three regimes of gravity in the nDGP model on different scales. 
On super-horizon scales, gravity is significantly influenced by 
5D effects. In this regime, we cannot ignore the time 
evolution of metric perturbations and the dynamical 
solutions should be obtained by solving the 5D equations 
of motion. The dynamical solutions have been obtained 
in the following two methods in the literature: a first derivation is 
obtained by the scaling ansatz in the sDGP \cite{Sawicki:2006jj}
and in the nDGP \cite{Song:2007wd}, 
and the other is found from the full 5D numerical simulations 
\cite{Cardoso:2007xc}. It has been shown 
that both approaches give identical results, and the solutions 
for the perturbations are shown to be insensitive to 
the initial conditions for the 5D metric perturbations. 

On sub-horizon scales, we can ignore the time dependence 
of the metric perturbations and the quasi-static 
approximations can be used \cite{Lue:2002sw, Koyama:2005kd}. Even on  scales smaller
than $r_c$, gravity is not described by general 
relativity due to an extra scalar degree of freedom 
introduced by the modification of gravity. 
In this regime, gravity can be described by 
a Brans-Dicke theory and the growth of structure 
becomes scale independent. 

\begin{figure}[htbp]
  \begin{center}
  \epsfysize=3.0truein
  \epsfxsize=3.0truein
    \epsffile{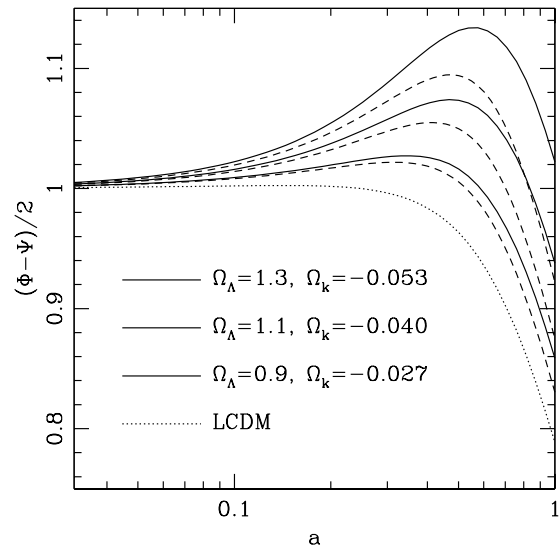}
    \caption{\footnotesize We plot the solutions of structure formation
of three nDGP models in the $1-\sigma$ contour of Fig.~\ref{fig:contour}, compared with the LCDM (dotted line).
Solid curves represent the quasi-static solutions of nDGP models
with different $\Omega_{\Lambda}$, and the dashed curve attached 
to each solid curve represents the dynamic solution of each nDGP model
at $k=10^{-3}$ Mpc$^{-1}$. 
}
\label{fig:GI}
\end{center}
\end{figure}

We use the Newtonian gauge
\begin{equation}
ds^2 =-(1+2\Psi) \, dt^2 + a(t)^2(1+2 \Phi) \, \delta_{ij} dx^i dx^j, 
\end{equation}
to describe the metric perturbations. 
Fig.~\ref{fig:GI} shows the behaviour of metric perturbations 
$\Phi_- \equiv (\Phi-\Psi)/2$ which determines the integrated 
Sachs-Wolfe (ISW) effect both for the dynamical 
solutions and scaling solutions, for the models of Table~\ref{tab:models}.
 In the literature, the spatial 
curvature was not introduced in the calculations, and thus
we derive the quasi-static solutions with curvature in Appendix A.

\begin{table}[htb]
\begin{center}
\begin{tabular}{| c | c | c | c | c |}
\hline
                     & LCDM   & nDGP 1  & nDGP 2  & nDGP 3   \\ 
\hline
$ \Omega_m $         & 0.30   & 0.32   & 0.34   & 0.37    \\
$ \Omega_b $         & 0.052  & 0.056  & 0.060  & 0.064   \\
$ \Omega_k $         & -0.014 & -0.027 & -0.040 & -0.053  \\
$ \Omega_{\Lambda} $ & 0.72   & 0.90   & 1.1    & 1.3     \\
$ H_0 $              & 66    & 63    & 61    & 59     \\
\hline
\end{tabular}
\caption{Details of the cosmological models used. The other parameters are for all models: scalar spectral index $ n_s = 0.95 $, optical depth $ \tau = 0.10 $, and amplitude of the primordial scalar perturbations $ A = 2.04 \cdot 10^{-9}$ at a pivot scale $ k = 0.05 $ Mpc$^{-1}$.}
\label{tab:models}
\end{center}
\end{table}

Finally, once the non-linearity of density perturbations becomes 
important, the theory approaches general relativity 
\cite{Lue:2002sw, Koyama:2007ih}. 
This transition to general relativity is crucial to satisfy the 
tight constraints from the solar system experiments \cite{Deffayet:2001uk,
Dvali:2002vf}, and
 will play 
a crucial role for weak lensing measures. On the 
other hand, for the ISW effect, we can safely ignore the 
non-linear physics. 

The dynamical solutions are relevant to the scales of the large scales 
CMB anisotropies. We have checked that the difference in the large 
scales CMB anisotropies from LCDM are small given the constraints 
from the geometrical tests because, due to the large cosmic variance, we do not 
expect that the CMB anisotropies on these scales can give strong constraints 
on the models. The quasi-static solutions are relevant to the scales of 
ISW-galaxy cross-correlations. In the next section, we will study how 
they can be used to break the degeneracy that arises from the geometrical 
tests. 

\section{ISW-galaxy correlations}
The gravitational potential well $\Phi_{-}$ is shallower in the nDGP model
than in the LCDM model due to the modification of gravity. This 
is the opposite from what happens in 
the self-accelerating models \cite{Song:2006jk} where
the gravitational potential well is deeper than in LCDM. 
The nDGP model predicts an earlier variation of the gravitational
potential than the LCDM model.
By cross-correlating galaxies at different redshifts with the CMB,
one can in principle trace the redshift history of the decay of the 
potential. Furthermore, the cross-correlation arises from the well understood
quasi-static (QS) regime of nDGP (solid curves in Fig.~\ref{fig:GI}).

\begin{figure*}[htb]
  \begin{center}
\includegraphics[angle=0,width=.9\linewidth]{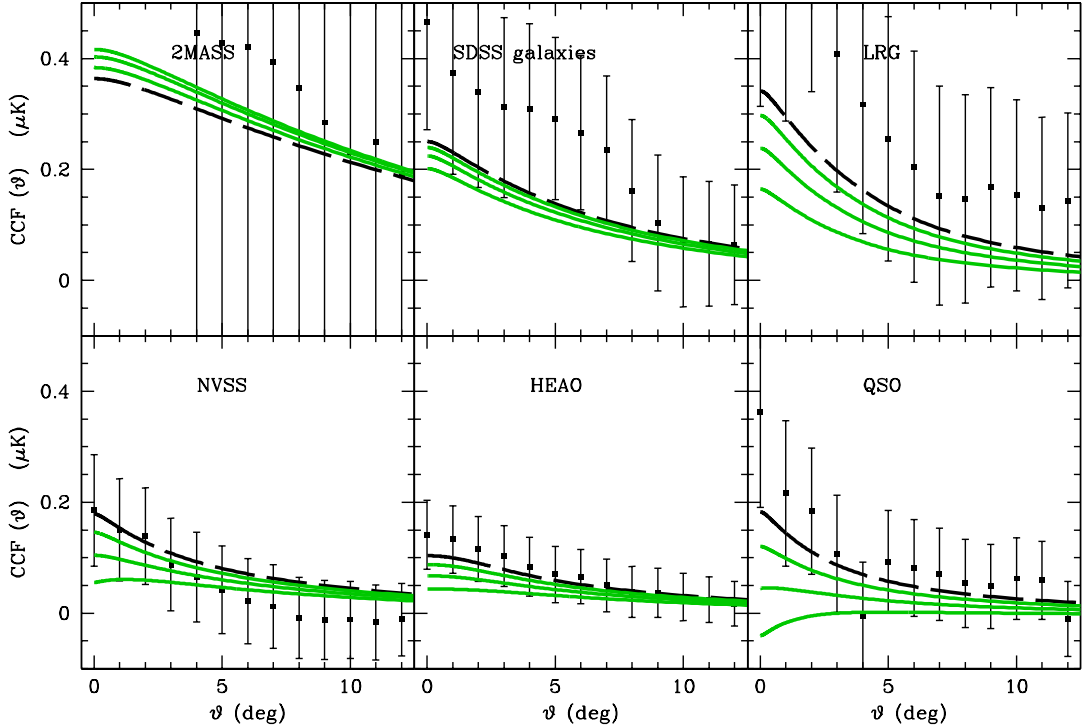}
    \caption{\footnotesize 
Measurement of the cross-correlation functions between six different galaxy data sets and the CMB,
 reproduced from \cite {Giannantonio:2008zi}. The curves show the theoretical predictions for the 
ISW-galaxy correlations at each redshift for the LCDM model (black, dashed) and the three nDGP
 models of Fig.~\ref{fig:GI} (green, solid), which describe the $1-\sigma $ region of the geometry test from Fig.~\ref{fig:contour}.
}
    \label{fig:constraints}
  \end{center}
\end{figure*}

The cross-power spectrum of the CMB and a set of galaxies $g_{i}$ is given by \cite {Pogosian:2004wa,Corasaniti:2005pq}
\ba
C^{{\rm I}g_{i}}_{\sc l}=4\pi \int \frac{{\rm d}k}{k}
I^{\rm I}_{\sc l}(k)I^{g_{i}}_{\sc l}(k){k^3P_{\Phi_-\Phi_-}(k,0)\over 2\pi^2},
\ea
where $P_{\Phi_-\Phi_-}(k,0)$ is the power spectrum of
$\Phi_{-}$ at the present time and
the kernel $I^{\rm I}_{\sc l}$ is
\ba
I^{\rm I}_{\sc l}(k)=\int {\rm d}z W^{\rm I}(k,z)
j_{\sc l}(kD) \left( {d D \over d r} \right)^{1/2}\,.
\ea
The window function is given by
\ba
W^{\rm I}(k,z)&=&
-\frac{2}{1+z}
\frac{\partial}{\partial \ln a}\left[\frac{\Phi_-(k,z)}{\Phi_-(k,0)}\right]\,,
\ea
where
the galaxy kernel $I^{g_{i}}$ is
\ba
I^{g_{i}}_{\sc l}(k)=\int {\rm d}z  W^{g_{i}}(k,z)
j_{\sc l}(kD) \left( {d D \over d r} \right)^{1/2}.
\ea
Under the QS approximation, the window function becomes
\ba
W^{g_{i}}(k,z)&=&\frac{2}{3\Omega_{m}}
\frac{k^2}{H_0^2}
\frac{n_{i}(z)b_{i}(z)}{1+z}\frac{\Phi_-(k,z)}{\Phi_-(k,0)}\,,
\ea
where $n_{i}(z)$ is the redshift distribution
of the galaxies normalised to $\int dz n_{i} = 1$
and $b_{i}(z)$ is the galaxy bias.

\begin{figure*}[t]
  \begin{center}
\includegraphics[angle=0,width=.8\linewidth]{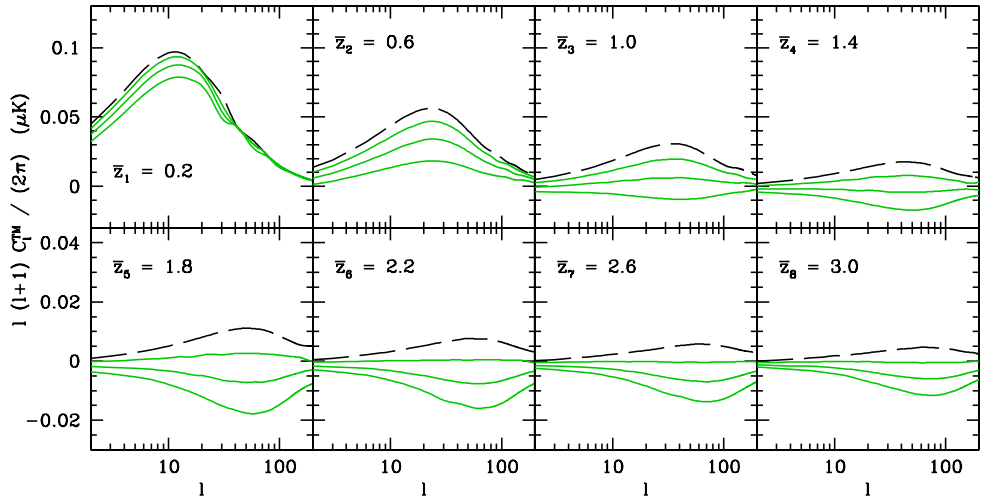}
    \caption{\footnotesize 
The theoretical predictions for the 
ISW-galaxy cross power spectra at each redshift for the LCDM model (black, dashed) and the three nDGP
 models of Fig.~\ref{fig:GI} (green, solid), which describe the $1-\sigma $ region of the geometry test from Fig.~\ref{fig:contour}.
}
    \label{fig:constraintsfut}
  \end{center}
\end{figure*}

First, we investigate the current status of the observations 
using the data set obtained in \cite {Giannantonio:2008zi}, which consists in the measurement of the ISW effect via the real space cross-correlation functions (CCF) between six different galaxy catalogues and the CMB. The redshift distributions of the different
catalogues are partly overlapping, but each data set $i$ is characterised by a median
redshift $ 0.1 \le \bar z_i \le 1.5$ around which each particular measured CCF is 
getting the biggest contribution. Thus, this analysis represents a first step 
towards an exploration of the redshift evolution of the potentials and, ultimately, 
of gravity itself as described by \cite{Hu:2004yd}.

We reproduce in Fig.~\ref{fig:constraints} the measured CCF for the six galaxy 
catalogues from \cite {Giannantonio:2008zi}, in order of increasing redshift: 2MASS (excluding the small scale contaminated data), the main galaxy sample from the SDSS, 
the SDSS Luminous Red Galaxies, NVSS, HEAO and the SDSS quasars, with the relative 
error bars which should be remembered are highly correlated. Looking at the 
theoretical curves in Fig.~\ref{fig:constraints}, we can see that
the nDGP models have a very different prediction from the LCDM for the CCF at high 
redshift. This is in agreement with their peculiar potential evolution: the rise 
in the potential $ \Phi_-$ at high redshift produces an expected negative CCF, while 
the following steeper decay leads to a positive CCF which becomes eventually higher 
than the LCDM one.

However, it is clear that these predictions represent a poor fit to the high redshift data. Remembering that all three nDGP models in Fig.~\ref{fig:constraints} are inside the $1\sigma$ region from the geometry test of Section II, we can qualitatively see that the ISW test will produce stricter constraints by noticing e.g. that the quasar CCF alone has a significance level of $ 2\sigma $, which means that at least two of the nDGP models will be excluded at above this level.

Then, we study the best possible constraints which can be obtained by this technique 
with future surveys. For definiteness, we assume that the galaxy sets come from a net 
galaxy distributions of
\ba
n_g(z)\propto z^2 e^{-(z/1.5)^2}\,,
\ea
where the normalisation is given by the LSST expectation
of 35 galaxies per $\rm arcmin^2$.
For the subsets of galaxies, we assume that this total distribution is
separated by photometric redshifts which have a Gaussian error
distribution with rms
$\sigma(z)=0.03(1+z)$.
The redshift distributions are then given by \cite{Hu:2004yd}
\ba
n_i(z)={A_{i}\over 2}n_g(z)
\left[ {\rm erfc}\left(\frac{z_{i-1}-z}{\sqrt{2}\sigma(z)}\right)
-{\rm erfc}\left(\frac{z_{i}-z}{\sqrt{2}\sigma(z)}\right)\right]\,,\nn
\ea
where erfc is the complementary error function and $A_{i}$ is determined
by the normalisation constraint.

We show in Fig.~\ref{fig:constraintsfut} the predicted cross power spectra obtained using this redshift tomography for the models of Table~\ref{tab:models}. The theoretical possibility to distinguish between them is given by the signal to noise ratio
\be
\left( \frac{S}{N} \right)^2 = \sum_l {\rm f_{sky}}(2l+1) \frac { [C^{Ig}_l]^2} {C_l^{gg} C_l^{TT} + [C_l^{Ig}]^2},
\ee
where $C_l^{TT}$ is the temperature power spectrum. This is summarised in Table~\ref{tab:SN}.

\begin{table}[htb]
\begin{center}
\begin{tabular}{| c | c | c | c | c |}
\hline
$ \bar z $ & LCDM   & nDGP 1  & nDGP 2  & nDGP 3   \\ 
\hline
0.2        & 2.8    & 2.9        & 2.6        &  2.2        \\
0.6        & 4.0    & 3.5        & 2.5        &  1.3        \\
1.0        & 3.4    & 2.2        & 0.68        &  1.1        \\
1.4        & 2.5    & 1.2        & 0.69        &  2.6        \\
1.8        & 1.9    & 0.52        & 1.3        &  3.2        \\
2.2        & 1.5    & 0.16        & 1.6        &  3.3        \\
2.6        & 2.4    & 0.18        & 1.6        &  3.1        \\
3.0        & 0.96    & 0.22        & 1.5        &  2.9        \\
\hline
\end{tabular}
\caption{Theoretical signal to noise ratio for the models of Table~\ref{tab:models} with ${\rm f_{sky}}=1$.}
\label{tab:SN}
\end{center}
\end{table}

Although the geometrical test is not able to easily break the degeneracy
between curvature and the screening effect,
the alternative consequence for the structure formation by the screening effect
is measurable from the ISW-galaxy cross-correlations.
The screening of the cosmological constant in nDGP2 and nDGP3 
becomes effective before the decay of the growth factor  
which occurs when the matter component becomes subdominant. This early screening 
enhances the growth factor which makes the potential $\Phi_-$ grow.
This generates anti-correlations in the ISW-galaxy cross-correlations 
at high redshifts, which leaves observable signatures as is 
shown in Fig.~\ref{fig:constraintsfut}.
From Table~\ref{tab:SN}, it is expected that this effect on the
structure formation can be observed at around 50$\%$ noise level for nDGP2 and 
25$\%$ noise level for nDGP3.
This is an illustration how we can break the degeneracy between 
curvature and the screening effect in the geometrical tests 
by using the structure formation tests.

\section{Conclusion}
In this paper, we studied the observational constraints on the normal branch 
DGP model in which a phantom-like behaviour occurs only with 
cold dark matter and a cosmological constant. 
The geometrical tests using the gold SN data set, CMB and the HST 
key project are not enough to rule out models in which gravity 
is significantly modified on cosmological scales. We then 
showed that the structure formation tests performed using the integrated 
Sachs-Wolfe (ISW) effect can break the degeneracy in the parameter 
space. 

The current measurements of the ISW effect obtained in \cite {Giannantonio:2008zi}
are indeed as competitive as the geometrical tests. This is due to 
the fact that, in the nDGP model, the cross-correlation with galaxies becomes negative
at high redshift due to the peculiar behaviour of the metric 
perturbations caused by the modification of gravity. This demonstrates 
that the structure formation tests are very promising tools to distinguish 
between general relativity and modified gravity models. We also showed that 
it is possible to track the evolution of the potentials by cross-correlating 
the ISW with galaxies at each redshift in future observations. 
It is very likely that in the future the ISW effect will provide one of the 
strongest constraints on the model. We will present the full likelihood 
analysis using the latest data sets in a forthcoming paper. 

\section*{Acknowledgements}
We would like to thank Robert Crittenden, Roy Maartens and Elisabetta Majerotto for useful discussion. YS and KK are supported by STFC.

\appendix
\section{Quasi-static solutions with curvature}
In the Gaussian normal coordinates, the 5D metric is 
given by \cite{Deffayet:2001uk}
\begin{equation}
ds^2 = dy^2 -n(y,t)^2 dt^2 + a(y,t)^2 \delta_{ij}dx^i dx^j,
\end{equation}
where
\begin{eqnarray}
a(y,t) &=& a(t) \left[1 - \left(H^2 -\frac{\Omega_k}{a_0^2} \right)^{\frac{1}{2}} y
\right], \\
n(y,t) &=&1 - (\dot{H}+H^2) \left(H^2 -\frac{\Omega_k}{a_0^2} \right)^{-\frac{1}{2}} y.
\end{eqnarray}
The extrinsic curvature of the brane is determined by the 
first derivative of the metric with respect to $y$ at the brane ($y=0$):
\begin{eqnarray}
\frac{a'}{a} & = & - \left(H^2 - \frac{\Omega_k}{a^2} \right)^{\frac{1}{2}}, \quad \\
\frac{n'}{n} & = & -\left(\dot{H} +H^2 \right)  \left(H^2 - \frac{\Omega_k}{a^2} \right)^{-\frac{1}{2}}.
\end{eqnarray}

Defining the comoving density perturbations
\begin{equation}
\rho \bigtriangleup = \delta \rho -3 Ha \delta q,
\end{equation}
the Poisson equation is obtained as
\begin{equation}
\frac{k^2}{a^2}\Phi = \frac{\kappa_4^2}{2} \left[
\frac{2 (a'/a)r_c}{2 (a'/a) r_c-1}  \right] \left[ \rho \bigtriangleup - 
\frac{\delta \rho_E-3 H a \delta q_E}{2 (a'/a) r_c} \right].
\label{poisson}
\end{equation}
The traceless part of the space-space component 
of the effective Einstein equations gives
\ba
-\frac{1}{a^2} && \!\!\!\!\!\!\!\!\!\!\!\left\{   
1 - \frac{1}{r_c \left[(a'/a)+(n'/n)\right]} \right\}
(\Phi+\Psi) \nonumber\\
&=&  - 
\frac{\kappa_4^2 \delta \pi_{E}}{r_c \left[(a'/a)+(n'/n)\right]}.
\label{anisotropy}
\ea
The Weyl density perturbations $\delta \rho_E$, $\delta q_E$ and 
$\delta \pi_E$ should be determined by the constraint equations 
\ba
\dot{\delta \rho}_E &+& 4 H \delta \rho_E - a^{-1} k^2 \delta q_E =0, \\
\dot{\delta q}_{E} &+& 4 H \delta q_E + a^{-1} 
\left(\frac{1}{3} \delta \rho_E - \frac{2}{3} k^2 \delta \pi_E \right)\nonumber\\
&=&-a^{-1} \frac{2}{3} r_c \left(\frac{n'}{n}-\frac{a'}{a} \right)
\left\{
-\frac{\rho \bigtriangleup}{2 (a'/a) r_c -1} \right. \nonumber\\
&+& \left.
\frac{\delta \rho_E -3 H a \delta q_E}{2 (a'/a)r_c-1} \right. \nonumber\\
 &+& \left.\frac{1}{r_c \left[(a'/a)+ (n'/n) \right]-1}
k^2 \delta \pi_E  \right\}. 
\label{EulerE}
\end{eqnarray}

The constraint equations are not closed and we need 
additional information by solving the 5D equation 
of motion. In the quasi-static limit, we can impose the 
condition on $\delta \rho_E$ and $\delta \pi_E$ from the bulk 
equation as \cite{Koyama:2005kd}
\begin{equation}
\delta \rho_E = 2 k^2 \delta \pi_E.
\end{equation}
Then the constraint equations give 
\begin{equation}
\delta \rho_E = 2 \left[    
\frac{-1+(a'/a)r_c + (n'/n)r_c}{-3 + 4 (a'/a) r_c + 2 (n'/n) r_c} \right]
\rho \Delta,
\end{equation}
and $\delta q_E=0$.
The Poisson equation and the traceless part of Einstein equations 
give 
\begin{eqnarray}
\frac{k^2}{a^2} \Phi &=& \frac{\kappa_4^2}{2} 
\left[1- \frac{1}{3 \beta(t)} \right] \rho \triangle,
\label{solphi}
\\
\frac{k^2}{a^2} \Psi &=& -\frac{\kappa_4^2}{2} 
\left[1 + \frac{1}{3 \beta(t)} \right] \rho \triangle,
\label{solpsi}
\end{eqnarray}
where
\begin{equation}
\beta(t) =1 -  \frac{2}{3} \left[2 \left(\frac{a'}{a}\right)+ 
\left(\frac{n'}{n}\right) \right] r_c,
\end{equation}
which can be written as 
\begin{equation}
\beta(t) = 1 + \; 2 H^2 r_c \left(H^2 - \frac{\Omega_k}{a^2} \right)^{-1/2}
\left[1+ \frac{\dot{H}}{3 H^2} - \frac{2}{3} \frac{\Omega_k}{a^2 H^2} \right].
\end{equation}

\bibliography{ms}

\begin{thebibliography}{27}
\expandafter\ifx\csname natexlab\endcsname\relax\def\natexlab#1{#1}\fi
\expandafter\ifx\csname bibnamefont\endcsname\relax
  \def\bibnamefont#1{#1}\fi
\expandafter\ifx\csname bibfnamefont\endcsname\relax
  \def\bibfnamefont#1{#1}\fi
\expandafter\ifx\csname citenamefont\endcsname\relax
  \def\citenamefont#1{#1}\fi
\expandafter\ifx\csname url\endcsname\relax
  \def\url#1{\texttt{#1}}\fi
\expandafter\ifx\csname urlprefix\endcsname\relax\def\urlprefix{URL }\fi
\providecommand{\bibinfo}[2]{#2}
\providecommand{\eprint}[2][]{\url{#2}}

\bibitem[{\citenamefont{Percival et~al.}(2007)}]{Percival:2007yw}
\bibinfo{author}{\bibfnamefont{W.~J.} \bibnamefont{Percival}}
  \bibnamefont{et~al.}, \bibinfo{journal}{Mon. Not. Roy. Astron. Soc.}
  \textbf{\bibinfo{volume}{381}}, \bibinfo{pages}{1053} (\bibinfo{year}{2007}),
  \eprint{arXiv:0705.3323 [astro-ph]}.

\bibitem[{\citenamefont{Giannantonio et~al.}(2008)}]{Giannantonio:2008zi}
\bibinfo{author}{\bibfnamefont{T.}~\bibnamefont{Giannantonio}}
  \bibnamefont{et~al.} (\bibinfo{year}{2008}), \eprint{arXiv:0801.4380
  [astro-ph]}.

\bibitem[{\citenamefont{Caldwell}(2002)}]{Caldwell:1999ew}
\bibinfo{author}{\bibfnamefont{R.~R.} \bibnamefont{Caldwell}},
  \bibinfo{journal}{Phys. Lett.} \textbf{\bibinfo{volume}{B545}},
  \bibinfo{pages}{23} (\bibinfo{year}{2002}), \eprint{astro-ph/9908168}.

\bibitem[{\citenamefont{Csaki et~al.}(2006)\citenamefont{Csaki, Kaloper, and
  Terning}}]{Csaki:2005vq}
\bibinfo{author}{\bibfnamefont{C.}~\bibnamefont{Csaki}},
  \bibinfo{author}{\bibfnamefont{N.}~\bibnamefont{Kaloper}}, \bibnamefont{and}
  \bibinfo{author}{\bibfnamefont{J.}~\bibnamefont{Terning}},
  \bibinfo{journal}{JCAP} \textbf{\bibinfo{volume}{0606}}, \bibinfo{pages}{022}
  (\bibinfo{year}{2006}), \eprint{astro-ph/0507148}.

\bibitem[{\citenamefont{Libanov et~al.}(2007)\citenamefont{Libanov, Rubakov,
  Papantonopoulos, Sami, and Tsujikawa}}]{Libanov:2007mq}
\bibinfo{author}{\bibfnamefont{M.}~\bibnamefont{Libanov}},
  \bibinfo{author}{\bibfnamefont{V.}~\bibnamefont{Rubakov}},
  \bibinfo{author}{\bibfnamefont{E.}~\bibnamefont{Papantonopoulos}},
  \bibinfo{author}{\bibfnamefont{M.}~\bibnamefont{Sami}}, \bibnamefont{and}
  \bibinfo{author}{\bibfnamefont{S.}~\bibnamefont{Tsujikawa}},
  \bibinfo{journal}{JCAP} \textbf{\bibinfo{volume}{0708}}, \bibinfo{pages}{010}
  (\bibinfo{year}{2007}), \eprint{arXiv:0704.1848 [hep-th]}.

\bibitem[{\citenamefont{Sahni and Shtanov}(2003)}]{Sahni:2002dx}
\bibinfo{author}{\bibfnamefont{V.}~\bibnamefont{Sahni}} \bibnamefont{and}
  \bibinfo{author}{\bibfnamefont{Y.}~\bibnamefont{Shtanov}},
  \bibinfo{journal}{JCAP} \textbf{\bibinfo{volume}{0311}}, \bibinfo{pages}{014}
  (\bibinfo{year}{2003}), \eprint{astro-ph/0202346}.

\bibitem[{\citenamefont{Lue and Starkman}(2004)}]{Lue:2004za}
\bibinfo{author}{\bibfnamefont{A.}~\bibnamefont{Lue}} \bibnamefont{and}
  \bibinfo{author}{\bibfnamefont{G.~D.} \bibnamefont{Starkman}},
  \bibinfo{journal}{Phys. Rev.} \textbf{\bibinfo{volume}{D70}},
  \bibinfo{pages}{101501} (\bibinfo{year}{2004}), \eprint{astro-ph/0408246}.

\bibitem[{\citenamefont{Dvali et~al.}(2000)\citenamefont{Dvali, Gabadadze, and
  Porrati}}]{Dvali:2000hr}
\bibinfo{author}{\bibfnamefont{G.~R.} \bibnamefont{Dvali}},
  \bibinfo{author}{\bibfnamefont{G.}~\bibnamefont{Gabadadze}},
  \bibnamefont{and} \bibinfo{author}{\bibfnamefont{M.}~\bibnamefont{Porrati}},
  \bibinfo{journal}{Phys. Lett.} \textbf{\bibinfo{volume}{B485}},
  \bibinfo{pages}{208} (\bibinfo{year}{2000}), \eprint{hep-th/0005016}.

\bibitem[{\citenamefont{Deffayet}(2001)}]{Deffayet:2000uy}
\bibinfo{author}{\bibfnamefont{C.}~\bibnamefont{Deffayet}},
  \bibinfo{journal}{Phys. Lett.} \textbf{\bibinfo{volume}{B502}},
  \bibinfo{pages}{199} (\bibinfo{year}{2001}), \eprint{hep-th/0010186}.

\bibitem[{\citenamefont{Koyama}(2007)}]{Koyama:2007za}
\bibinfo{author}{\bibfnamefont{K.}~\bibnamefont{Koyama}}
  (\bibinfo{year}{2007}), \eprint{arXiv:0709.2399 [hep-th]}.

\bibitem[{\citenamefont{Lazkoz et~al.}(2006)\citenamefont{Lazkoz, Maartens, and
  Majerotto}}]{Lazkoz:2006gp}
\bibinfo{author}{\bibfnamefont{R.}~\bibnamefont{Lazkoz}},
  \bibinfo{author}{\bibfnamefont{R.}~\bibnamefont{Maartens}}, \bibnamefont{and}
  \bibinfo{author}{\bibfnamefont{E.}~\bibnamefont{Majerotto}},
  \bibinfo{journal}{Phys. Rev.} \textbf{\bibinfo{volume}{D74}},
  \bibinfo{pages}{083510} (\bibinfo{year}{2006}), \eprint{astro-ph/0605701}.

\bibitem[{\citenamefont{Lazkoz and Majerotto}(2007)}]{Lazkoz:2007zk}
\bibinfo{author}{\bibfnamefont{R.}~\bibnamefont{Lazkoz}} \bibnamefont{and}
  \bibinfo{author}{\bibfnamefont{E.}~\bibnamefont{Majerotto}},
  \bibinfo{journal}{JCAP} \textbf{\bibinfo{volume}{0707}}, \bibinfo{pages}{015}
  (\bibinfo{year}{2007}), \eprint{arXiv:0704.2606 [astro-ph]}.

\bibitem[{\citenamefont{Riess et~al.}(2004)}]{riess04}
\bibinfo{author}{\bibfnamefont{A.~G.} \bibnamefont{Riess}} \bibnamefont{et~al.}
  (\bibinfo{collaboration}{Supernova Search Team}),
  \bibinfo{journal}{Astrophys. J.} \textbf{\bibinfo{volume}{607}},
  \bibinfo{pages}{665} (\bibinfo{year}{2004}), \eprint{astro-ph/0402512}.

\bibitem[{\citenamefont{Spergel et~al.}(2003)}]{spergel03}
\bibinfo{author}{\bibfnamefont{D.~N.} \bibnamefont{Spergel}}
  \bibnamefont{et~al.} (\bibinfo{collaboration}{WMAP}),
  \bibinfo{journal}{Astrophys. J. Suppl.} \textbf{\bibinfo{volume}{148}},
  \bibinfo{pages}{175} (\bibinfo{year}{2003}), \eprint{astro-ph/0302209}.

\bibitem[{\citenamefont{Freedman et~al.}(2001)}]{Freedman:2000cf}
\bibinfo{author}{\bibfnamefont{W.~L.} \bibnamefont{Freedman}}
  \bibnamefont{et~al.} (\bibinfo{collaboration}{HST}),
  \bibinfo{journal}{Astrophys. J.} \textbf{\bibinfo{volume}{553}},
  \bibinfo{pages}{47} (\bibinfo{year}{2001}), \eprint{astro-ph/0012376}.

\bibitem[{\citenamefont{Sawicki et~al.}(2007)\citenamefont{Sawicki, Song, and
  Hu}}]{Sawicki:2006jj}
\bibinfo{author}{\bibfnamefont{I.}~\bibnamefont{Sawicki}},
  \bibinfo{author}{\bibfnamefont{Y.-S.} \bibnamefont{Song}}, \bibnamefont{and}
  \bibinfo{author}{\bibfnamefont{W.}~\bibnamefont{Hu}}, \bibinfo{journal}{Phys.
  Rev.} \textbf{\bibinfo{volume}{D75}}, \bibinfo{pages}{064002}
  (\bibinfo{year}{2007}), \eprint{astro-ph/0606285}.

\bibitem[{\citenamefont{Song}(2007)}]{Song:2007wd}
\bibinfo{author}{\bibfnamefont{Y.-S.} \bibnamefont{Song}}
  (\bibinfo{year}{2007}), \eprint{arXiv:0711.2513 [astro-ph]}.

\bibitem[{\citenamefont{Cardoso et~al.}(2007)\citenamefont{Cardoso, Koyama,
  Seahra, and Silva}}]{Cardoso:2007xc}
\bibinfo{author}{\bibfnamefont{A.}~\bibnamefont{Cardoso}},
  \bibinfo{author}{\bibfnamefont{K.}~\bibnamefont{Koyama}},
  \bibinfo{author}{\bibfnamefont{S.~S.} \bibnamefont{Seahra}},
  \bibnamefont{and} \bibinfo{author}{\bibfnamefont{F.~P.} \bibnamefont{Silva}}
  (\bibinfo{year}{2007}), \eprint{arXiv:0711.2563 [astro-ph]}.

\bibitem[{\citenamefont{Lue and Starkman}(2003)}]{Lue:2002sw}
\bibinfo{author}{\bibfnamefont{A.}~\bibnamefont{Lue}} \bibnamefont{and}
  \bibinfo{author}{\bibfnamefont{G.}~\bibnamefont{Starkman}},
  \bibinfo{journal}{Phys. Rev.} \textbf{\bibinfo{volume}{D67}},
  \bibinfo{pages}{064002} (\bibinfo{year}{2003}), \eprint{astro-ph/0212083}.

\bibitem[{\citenamefont{Koyama and Maartens}(2006)}]{Koyama:2005kd}
\bibinfo{author}{\bibfnamefont{K.}~\bibnamefont{Koyama}} \bibnamefont{and}
  \bibinfo{author}{\bibfnamefont{R.}~\bibnamefont{Maartens}},
  \bibinfo{journal}{JCAP} \textbf{\bibinfo{volume}{0601}}, \bibinfo{pages}{016}
  (\bibinfo{year}{2006}), \eprint{astro-ph/0511634}.

\bibitem[{\citenamefont{Koyama and Silva}(2007)}]{Koyama:2007ih}
\bibinfo{author}{\bibfnamefont{K.}~\bibnamefont{Koyama}} \bibnamefont{and}
  \bibinfo{author}{\bibfnamefont{F.~P.} \bibnamefont{Silva}},
  \bibinfo{journal}{Phys. Rev.} \textbf{\bibinfo{volume}{D75}},
  \bibinfo{pages}{084040} (\bibinfo{year}{2007}), \eprint{hep-th/0702169}.

\bibitem[{\citenamefont{Deffayet et~al.}(2002)\citenamefont{Deffayet, Dvali,
  Gabadadze, and Vainshtein}}]{Deffayet:2001uk}
\bibinfo{author}{\bibfnamefont{C.}~\bibnamefont{Deffayet}},
  \bibinfo{author}{\bibfnamefont{G.~R.} \bibnamefont{Dvali}},
  \bibinfo{author}{\bibfnamefont{G.}~\bibnamefont{Gabadadze}},
  \bibnamefont{and} \bibinfo{author}{\bibfnamefont{A.~I.}
  \bibnamefont{Vainshtein}}, \bibinfo{journal}{Phys. Rev.}
  \textbf{\bibinfo{volume}{D65}}, \bibinfo{pages}{044026}
  (\bibinfo{year}{2002}), \eprint{hep-th/0106001}.

\bibitem[{\citenamefont{Dvali et~al.}(2003)\citenamefont{Dvali, Gruzinov, and
  Zaldarriaga}}]{Dvali:2002vf}
\bibinfo{author}{\bibfnamefont{G.}~\bibnamefont{Dvali}},
  \bibinfo{author}{\bibfnamefont{A.}~\bibnamefont{Gruzinov}}, \bibnamefont{and}
  \bibinfo{author}{\bibfnamefont{M.}~\bibnamefont{Zaldarriaga}},
  \bibinfo{journal}{Phys. Rev.} \textbf{\bibinfo{volume}{D68}},
  \bibinfo{pages}{024012} (\bibinfo{year}{2003}), \eprint{hep-ph/0212069}.

\bibitem[{\citenamefont{Song et~al.}(2007)\citenamefont{Song, Sawicki, and
  Hu}}]{Song:2006jk}
\bibinfo{author}{\bibfnamefont{Y.-S.} \bibnamefont{Song}},
  \bibinfo{author}{\bibfnamefont{I.}~\bibnamefont{Sawicki}}, \bibnamefont{and}
  \bibinfo{author}{\bibfnamefont{W.}~\bibnamefont{Hu}}, \bibinfo{journal}{Phys.
  Rev.} \textbf{\bibinfo{volume}{D75}}, \bibinfo{pages}{064003}
  (\bibinfo{year}{2007}), \eprint{astro-ph/0606286}.

\bibitem[{\citenamefont{Pogosian}(2005)}]{Pogosian:2004wa}
\bibinfo{author}{\bibfnamefont{L.}~\bibnamefont{Pogosian}},
  \bibinfo{journal}{JCAP} \textbf{\bibinfo{volume}{0504}}, \bibinfo{pages}{015}
  (\bibinfo{year}{2005}), \eprint{astro-ph/0409059}.

\bibitem[{\citenamefont{Corasaniti et~al.}(2005)\citenamefont{Corasaniti,
  Giannantonio, and Melchiorri}}]{Corasaniti:2005pq}
\bibinfo{author}{\bibfnamefont{P.-S.} \bibnamefont{Corasaniti}},
  \bibinfo{author}{\bibfnamefont{T.}~\bibnamefont{Giannantonio}},
  \bibnamefont{and}
  \bibinfo{author}{\bibfnamefont{A.}~\bibnamefont{Melchiorri}},
  \bibinfo{journal}{Phys. Rev.} \textbf{\bibinfo{volume}{D71}},
  \bibinfo{pages}{123521} (\bibinfo{year}{2005}), \eprint{astro-ph/0504115}.

\bibitem[{\citenamefont{Hu and Scranton}(2004)}]{Hu:2004yd}
\bibinfo{author}{\bibfnamefont{W.}~\bibnamefont{Hu}} \bibnamefont{and}
  \bibinfo{author}{\bibfnamefont{R.}~\bibnamefont{Scranton}},
  \bibinfo{journal}{Phys. Rev.} \textbf{\bibinfo{volume}{D70}},
  \bibinfo{pages}{123002} (\bibinfo{year}{2004}), \eprint{astro-ph/0408456}.

\end{thebibliography}

\end{document}